\documentclass[twocolumn,pre,superscriptaddress,floatfix]{revtex4}
\usepackage{graphicx}
\newcommand{\kB}{k_{\mathrm{B}}}
\newcommand{\kT}{\kB T}
\newcommand{\lB}{l_{\mathrm{B}}}
\newcommand{\dens}{\rho}
\newcommand{\rhoback}{\dens_{\mathrm{back}}}
\newcommand{\rhop}{\dens_+}
\newcommand{\rhom}{\dens_-}
\newcommand{\rhosalt}{\dens_s}
\newcommand{\rhoprot}{\dens_p}
\newcommand{\rhopm}{\dens_\pm}

\newcommand{\chempot}{\mu}

\newcommand{\mupm}{\chempot_\pm}
\newcommand{\kap}{\kappa}
\newcommand{\kappas}{\kap_s}
\newcommand{\phid}{\phi_{\mathrm{D}}}
\newcommand{\surftens}{\gamma}
\newcommand{\dsurftens}{\Delta\surftens_{\mathrm{el}}}
\newcommand{\surfinf}{\surftens_0}
\newcommand{\welldepth}{\epsilon}
\newcommand{\length}{l}
\newcommand{\Q}{Q}
\newcommand{\df}{W}
\newcommand{\nm}{\mathrm{nm}}
\newcommand{\milliN}{\mathrm{mN}}
\newcommand{\microN}{\mu\mathrm{N}}
\newcommand{\metre}{\mathrm{m}}
\newcommand{\pH}{p\mathrm{H}}
\newcommand{\Molar}{\mathrm{M}}
\newcommand{\mC}{\mathrm{mC}}
\newcommand{\PB}{Poisson-Boltzmann}
\newcommand{\etal}{\emph{et al}}
\newcommand{\eqref}[1]{(\ref{#1})}

\begin{document}

\title{On the electrical double layer contribution to the 
interfacial tension of protein crystals}

\author{R. P. Sear}

\affiliation{Department of Physics, University of Surrey, Guildford,
Surrey GU2 7XH, United Kingdom}

\author{P. B. Warren}

\affiliation{Unilever Research and Development Port Sunlight, 
Bebington, Wirral CH63 3JW, United Kingdom}

\date{5 June 2002 --- PREPRINT}

\begin{abstract}
We study the electrical double layer at the interface between a
protein crystal and a salt solution or a dilute solution of protein,
and estimate the double layer's contribution to the interfacial
tension of this interface.  This contribution is negative and
decreases in magnitude with increasing salt concentration.  We also
consider briefly the interaction between a pair of protein surfaces.
\end{abstract}

%\pacs{Valid PACS}

\maketitle

\section{Introduction}
Protein crystallisation is not only of immense practical importance,
but also attracts interest as a phenomenon in its own right.  The
conditions under which crystallisation readily occurs are elusive and
vary unpredictably, and a number of hypotheses to explain this have
been put forward \cite{Poon,tWF,Sear,Sear99,Sear02,Piazza}.  The
process of crystallisation, as with other first-order phase
transitions, begins with nucleation, where a microscopic nucleus of
the crystalline phase first forms.  The elementary theory of
nucleation, generally called classical nucleation theory
\cite{Debenedetti}, predicts a rate of nucleation which varies as the
exponential of minus the cube of the interfacial tension and hence is
very sensitive to the magnitude of the interfacial tension.  In the
present paper, we extend earlier work of one of us on the effect of
salt on the bulk phase behaviour \cite{PBW}, to calculate the effect
of salt on the interfacial tension and electrical structure at the
surface of a protein crystal. Our calculation can be compared to the
recent work by Haas and Drenth who also develop a theory for the
interfacial tension of protein crystals \cite{HD}. Their theory is of
the Cahn-Hilliard type and does not treat electrostatic effects
explicitly.

It is a general consequence of the long range nature of the Coulomb
law that all bulk phases have to be electrically neutral and space
charge effects are confined to interfacial regions.  The consequences
of charge neutrality for membrane equilibrium were explored by
Frederick Donnan as long ago as 1911 \cite{Donnan}.  Much more
recently, the electrical structure of the interface in an electrolyte
near the critical point was studied by Nabutovskii and Nemov
\cite{NN}. In a crystal of charged protein molecules, the counterion
concentration in the interior of the crystal will be higher than in
the salt solution outside the crystal and the coion concentration will
be lower inside the crystal. The higher counterion concentration and
lower coion concentration will be such that the net charge density in
the crystal vanishes.  If one can ignore specific ion interactions and
the effects of non-ideality, the ion concentrations can be determined
using Donnan's 1911 theory, treating the protein-solution interface as
a membrane. This was done in Ref.~\cite{PBW}.

\begin{figure}
\begin{center}
\includegraphics{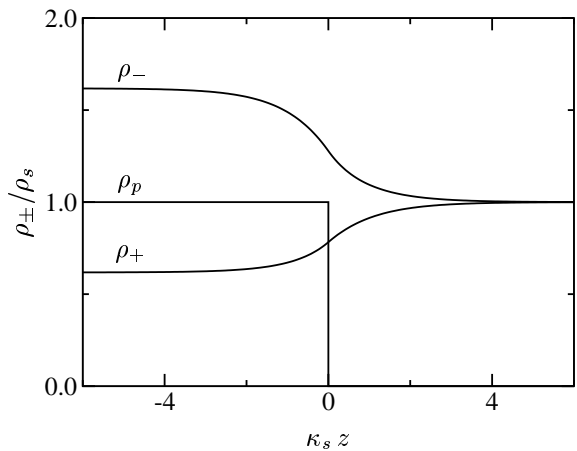}
\end{center}
\caption[?]{Ion density profiles for jellium model of a protein
crystal-solution interface.  The density profiles are computed from
the numerical solution to the Poisson-Boltzmann equation, and have
been normalised by the salt concentration $\rhosalt$.  In this
example, the protein charge density $\rhoprot$ (assumed positive) was
chosen equal to $\rhosalt$. The densities of counter and coions are
$\rhom$ and $\rhop$, respectively.  Distance is normalised by the
Debye screening length defined in
Eq.~\eqref{kseq}.\label{fig-profiles}}
\end{figure}

In the vicinity of the interface, the ion densities must vary smoothly
between the interior of the crystal and the exterior salt solution.
In particular, as shown in Fig.~\ref{fig-profiles}, one would expect
counterions to spill out into the salt solution and coions to
similarly spill into the protein, leading to the formation of an
electrical double layer across the interface.  The space charge gives
rise to an electric field and a jump in the mean electrostatic
potential between the salt solution and the protein crystal.  This
potential difference is sometimes termed the Donnan potential, or the
Galvani potential \cite{Fisher}.  In Donnan's theory, this potential
jump is self-consistently responsible for the change in ion
concentrations in the bulk phases.

In the theory of metals, a closely analogous phenomenon occurs at a
metal surface where the electron density spills out into vacuum.  The
resulting double layer often makes an important contribution to the
work function of metals, and was one of the early targets for density
functional theory \cite{Smith}.  In the metal case there is only one
density, namely the electron density, which vanishes in the vacuum.
The electron density in the interior is very high and the electrons
form a highly degenerate Fermi liquid, so that careful attention has
to be paid to exchange and correlation effects.  In contrast, in the
protein case one has two densities, for counterions and coions, which
approach a common value in the salt solution outside the crystal.
Moreover, for a 1:1 electrolyte in a high dielectric solvent,
correlation effects are not that important and for simplicity in our
calculation we ignore them.

In the next section we determine the potential profile at the
crystal-solution interface, and calculate the associated free energy
which represents the contribution of the electrical structure to the
interfacial tension.  The third section contains results of these
calculations along with approximate analytic expressions for the
magnitude of this contribution. The last section is a conclusion.

\section{Calculation}
We consider the case where we have a dense protein crystal coexisting
with a very dilute protein solution, sufficiently dilute that the few
protein molecules in it can be neglected and it treated as simply a
salt solution.  Rather than considering a specific protein crystal, we
take over the jellium concept from the theory of metals to make a
general estimate of the effect of salt on the interfacial tension.  We
replace the detailed charge density due to the protein molecules by a
uniform background charge density, cut off abruptly at the interface.
Thus our model of the bare protein crystal comprises a uniformly
charged half-space, with a charge density
\begin{equation}
\rhoback(z)=\left\{
\begin{array}{ll}
\rhoprot&(z<0)\\
0&(z>0)
\end{array}\right.
\label{back}
\end{equation}
where $\rhoprot$ is the mean charge density in the protein crystal.
For definiteness we take the protein to be positively charged so
the counterions are negative and the coions are positive; both are
monovalent. The discontinuity in background charge density at $z=0$
is dressed by the counterion and coion densities $\rhom(z)$ and
$\rhop(z)$.  These satisfy $\rhopm\to\rhosalt$ as $z\to\infty$ where
$\rhosalt$ is the salt concentration, and $(\rhom-\rhop)\to\rhoprot$
as $z\to-\infty$.

For notational convenience, we work in units where $e=\kT=1$.  In
these units, the Coulomb potential energy $U$ between a pair of
elementary charges separated by $r$ is $U=\lB/r$, where $\lB$ is the
Bjerrum length, equal to $0.72\,\mathrm{nm}$ in water at room
temperature ($\lB=e^2/4\pi\epsilon\kT$).  We assume a constant value
of $\lB$, and ignore dielectric effects.

Below, we will find that in the limit of very large salt
concentrations the potential is uniform everywhere, in the protein
crystal and in solution, and so can be set equal to zero. In this
limit the only relevant contributions to the free energy are
independent of salt concentration. For the purposes of considering the
effect of reducing the salt concentration, we assume that when the
concentration is reduced the protein crystal remains unchanged and
interacts with the salt ions solely through a mean electrostatic
potential, which we will calculate. We can then approximate the
protein crystal by a uniform background charge density,
Eq.~\eqref{back}, and ignore the rest of the protein interactions
until we come to estimate their contribution to the interfacial
tension. These assumptions are reasonable if the interactions of the
protein molecules are strong enough to form a dense, relatively rigid,
crystal.

We will use a grand potential, $\Omega_{el}$, which contains only
ideal solution terms for the ions and the associated electric field at
the interface.  We use a grand potential because as usual the
calculation of the interfacial tension is easiest at fixed chemical
potential not density. The grand potential $\Omega_{el}$, is,
\begin{equation}
\Omega_{el}=\int_{-\infty}^\infty\!dz\,\omega(z),\quad
\omega=\sum_{i=\pm}\dens_i(\log\frac{\dens_i}{\rhosalt}-1)
+\frac{E^2}{8\pi\lB}.\label{oeq}
\end{equation}
The first terms in $\omega$ are the ideal solution terms (the ions
share a common chemical potential $\mupm=\log\rhosalt$).  The last
term is the electrostatic energy, wherein $E=-d\phi/dz$ is the
electric field strength corresponding to an electrostatic potential
$\phi$ which satisfies the Poisson equation,
\begin{equation}
\frac{d^2\phi}{dz^2} +
4\pi\lB(\rhop - \rhom + \rhoback) = 0.
\end{equation}
A correlation term $\Omega_{\mathrm{corr}}[\rhopm(z)]$ could also be
included in Eq.~\eqref{oeq} but, as already mentioned, for the
purposes of the present calculation we omit correlation effects and
work within the simple mean field theory.

The variational principle $\delta\Omega_{el}/\delta[\rhopm(z)]=0$
applied in this problem yields $\rhopm(z)=\rhosalt\exp[\mp\phi]$.  The
electrostatic potential then satisfies the \PB\ equation,
\begin{equation}
\frac{d^2\phi}{dz^2}-\kappas^2\sinh\phi = 
\left\{\begin{array}{ll}
-4\pi\lB\rhoprot&(z<0)\\
0&(z>0)
\end{array}\right.\label{pbeq}
\end{equation}
where 
\begin{equation}
\kappas^2=8\pi\lB\rhosalt.\label{kseq}
\end{equation}
Equation \eqref{pbeq} can be integrated once with respect to $\phi$ to
obtain
\begin{equation}
\left(\frac{d\phi}{dz}\right)^2=
\left\{\begin{array}{ll}
2\kappas^2\left(\cosh\phi-\cosh\phid\right)& \\
\qquad\quad-8\pi\lB\rhoprot\left(\phi-\phid\right)
&(z<0)\\
&\\
2\kappas^2\left(\cosh\phi-1\right)&(z>0)
\end{array}\right.\label{pbeqint}
\end{equation}
where $\phid$ is the Donnan potential, the potential $\phi$ in the
bulk of the protein crystal, so $\phi\to\phid$ as $z\to-\infty$.
The potential in the bulk of the solution is 0, $\phi\to0$
as $z\to\infty$.

For $z>0$ Eq.~\eqref{pbeqint} can be integrated analytically
\cite{Israelachvili} to get $\phi = 2\log[(1+Ce^{-\kappas
z})/(1-Ce^{-\kappas z})]$ where $C = \tanh(\phi_0/4)$ and $\phi_0 =
\phi(0)$ is the value of the potential at the interface itself.  For
$z<0$, it is not possible to integrate analytically, it must be done
numerically.  Now, the charge densities for the co- and counterions in
the bulk of the protein crystal (i.e., $z\to-\infty$) are
\begin{equation}
\rho_{\pm}(-\infty)=\rhosalt\exp(\mp\phid).
\label{rhopm}
\end{equation}
Charge neutrality imposes
\begin{equation}
\rho_{+}(-\infty)+\rho_p=\rho_{-}(-\infty).
\end{equation}
Combining these 2 equations yields an equation for the Donnan
potential
\begin{equation}
\sinh\phid=\frac{\rhoprot}{2\rhosalt}\label{doneq}.
\end{equation}
and using this equation and Eq.~\eqref{kseq} in Eq.~\eqref{pbeqint} we
have
\begin{equation}
\begin{array}{r}
({d\phi}/{dz})^2=2\kappas^2
[\cosh\phi-\cosh\phid\qquad\\
-(\phi-\phid)\sinh\phid]
\end{array}\quad(z<0).
\label{odeeq}
\end{equation}
Matching $d\phi/dz$ across $z=0$ using Eq.~\eqref{pbeqint} for $z>0$
and Eq.~\eqref{odeeq} for $z<0$, gives
\begin{equation}
\phi_0=\phid+({1-\cosh\phid})/{\sinh\phid}.
\label{phi0}
\end{equation}
This not only fixes the complete solution for $z>0$, but can also be
used as the starting point for a numerical integration of
Eq.~\eqref{odeeq} into the $z<0$ half-space.  Note that if
Eq.~\eqref{odeeq} is used to determine $d\phi/dz$, the negative root
should be taken; also the term $[\dots]$ in this equation has a
geometric interpretation as the distance between $\cosh\phi$ and its
tangent at $\phi=\phid$, and thus is always positive.

\begin{figure}
\begin{center}
\includegraphics{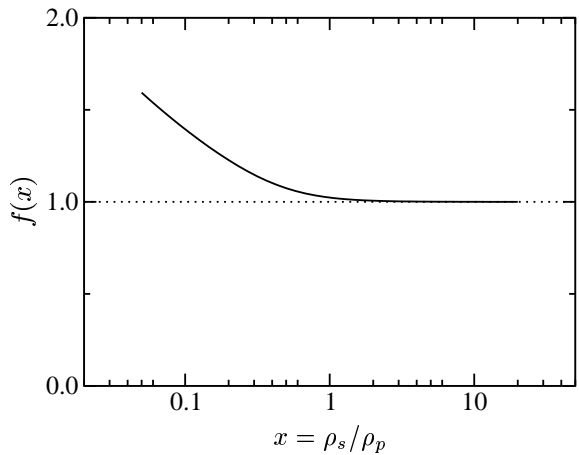}
\end{center}
\caption[?]{Numerical factor in the interfacial tension,
Eq.~\eqref{dsgeneq}, as determined from the full numerical solution
to the Poisson-Boltzmann equation.\label{fig-function}}
\end{figure}

The benefit of starting with an expression for the grand potential is
that the electrical contribution to the interfacial tension,
$\dsurftens$, is easily calculated. The
interfacial tension is the difference between the actual
grand potential per unit area of the interface and that it would
have if each of the 2 phases continued unperturbed
right up to a sharp dividing line between them.
We therefore have to calculate the grand potential then
subtract the grand potentials for the bulk protein and
solution states, thus
\begin{equation}
\dsurftens=
\int_{-\infty}^\infty\!dz\,
[\omega(z) - \theta(-z)\omega(-\infty)
 - \theta(z)\omega(+\infty)]\label{dseq1}
\end{equation}
where $\theta(z)$ is the Heaviside step-function, and
$\omega(\pm\infty)$ are the limiting values of the grand potential
density at $z\to\pm\infty$. They may be obtained from the definition
of $\omega$ in Eq.~\eqref{oeq} and Eq.~\eqref{rhopm} for the crystal,
\begin{equation}
\begin{array}{l}
\omega(-\infty)=-2\rhosalt\left(\cosh\phid
 - \phid\sinh\phid\right)\\
\omega(+\infty)=-2\rhosalt
\end{array}
\label{omegas}
\end{equation}
noting that in the bulk phases the electric field is zero.  Now, using
Eq.~\eqref{oeq} together with $\rho_{\pm}=\rho_s\exp(\mp\phi)$ and
$E(z)=-d\phi/dz$ from Eqs.~\eqref{pbeqint} and \eqref{odeeq}, we can
write $\omega(z)$ as a function of $\phi$. Using this together with
Eq.~\eqref{omegas} we can then write Eq.~\eqref{dseq1} as
\begin{equation}
\dsurftens=2\rhosalt\int_{-\infty}^\infty\!dz\,
\phi[\sinh\phi-\theta(-z)\sinh\phid].\label{dseq}
\end{equation}

\section{Results}\label{sec:results}
Before turning to the results of our full calculations, we discuss the
high salt limit.  If $\rhosalt\gg\rhoprot$ then $\phi\ll1$ everywhere
and the \PB\ equation (Eq.~\eqref{pbeqint} for $z>0$ and
Eq.~\eqref{odeeq} for $z<0$) can be linearised. The Donnan potential
is $\phid = {\rhoprot} / {2\rhosalt}$, the potential at $z=0$,
Eq.~\eqref{phi0}, becomes $\phi_0=\phid/2$, and the solution to the
linearised \PB\ equation is
\begin{equation}
\phi=\left\{\begin{array}{ll}
(\phid/2)\left[2-\exp(\kappas z)\right]&(z<0)\\
(\phid/2)\exp(-\kappas z)  &(z>0)
\end{array}
\right. .
\end{equation}
In Eq.~\eqref{dseq} these yield for the interfacial tension
\begin{equation}
\dsurftens =-\frac{\rhosalt\phid^2}{2\kappas}
\quad(\rhosalt\gg\rhoprot).\label{dshseq}
\end{equation}

This already shows the main features of the full solution, being
\emph{negative} and of decreasing magnitude as the salt concentration
increases.  The fact that the result should be negative can be seen
directly from Eq.~\eqref{dseq1}.  If the ion density profiles were
such that they remained constant right up to the interface where they
had a sharp jump between the bulk values, then Eq.~\eqref{dseq1} would
give $\dsurftens=0$. The fact that the ion density profiles are
relaxed compared to such a trial density profile implies that
$\dsurftens < 0$.  Note that the electrostatic energy is always
positive though (it is the last term in Eq.~\eqref{oeq}).

For the general case, we note that $\kappa_s^{-1}$ is the only length
scale that enters into the \PB\ equation, thus dimensional analysis
allows us to write
\begin{equation}
\dsurftens =-\frac{\rhosalt\phid^2}{2\kappas}
\,f(\frac{\rhosalt}{\rhoprot})\label{dsgeneq}
\end{equation}
where $f(x)$ is a numerical factor which amends the high salt scaling
limit and depends only on the ratio $x=\rhosalt/\rhoprot$.  We have
implemented a numerical scheme to integrate Eqs.~\eqref{odeeq} and
\eqref{dseq}, and determine $f(x)$.  The result is shown in
Fig.~\ref{fig-function}.  The function obviously obeys $f\to1$ as
$\rhosalt/\rhoprot\to\infty$, but the interesting thing is that $f$
does not deviate greatly from unity even at much lower salt
concentrations, for example $f\approx1.4$ at $\rhosalt / \rhoprot
\approx 0.1$.  This indicates that the high-salt scaling limit
Eq.~\eqref{dshseq} is not a bad approximation even at much lower salt
concentrations, provided that we use the exact expression for the
Donnan potential given in Eq.~\eqref{doneq}.

\begin{figure}
\begin{center}
\includegraphics{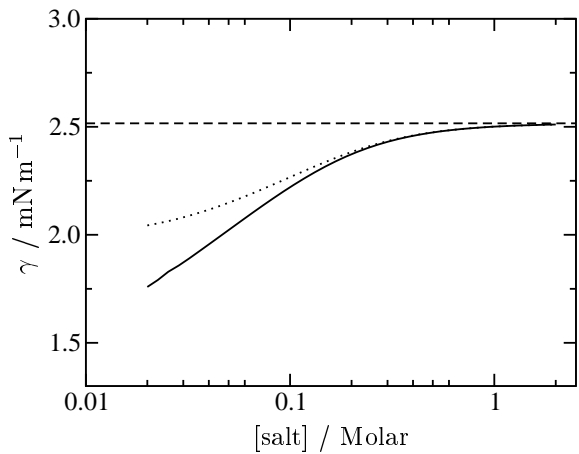}
\end{center}
\caption[?]{Predicted interfacial tension for lysozyme crystals, based
on the present work and previous models \cite{Sear,PBW}.  The dashed
line is the high-salt limit $\surfinf$ estimated in the text.  The
solid line incorporates the correction in Fig.~\ref{fig-function}.
The dotted line is the result if $f=1$ is used in
Eq.~\eqref{dsgeneq}.\label{fig-lysotens}}
\end{figure}

We can further approximate the contribution to the surface potential
by using the high salt expression for $\phid$, which is obtained by
linearising Eq.~\eqref{doneq}, and is $\phid=\rhoprot/(2\rhosalt)$.
With this approximation we obtain $\dsurftens = - \rhoprot^2 /
(8\rhosalt\kappas)$.  At high salt the contribution of the ions to the
thermodynamic potential of the protein crystal is again obtained by
linearisation and is given by $\omega(-\infty) = - 2\rhosalt +
\rhoprot^2 / (4\rhosalt)$.  The first term is simply
$\omega(+\infty)$, it is linear in $\rhosalt$ and has no effect, the
second term is proportional to the square of the protein density and
so is a, positive, contribution to the second virial coefficient. This
has been noted before, see Refs.~\cite{PBW,PoonEgel}. So, the
contribution to the interfacial tension is of order of the
contribution to the free energy density times the Debye length
$\kappas^{-1}$, which defines the thickness of the interface of
course.  This is what we might have expected simply on the basis of
dimensional analysis, although dimensional analysis would not of
course have told us that $\dsurftens$ is negative.

Our proteins are charged colloids and at least at high salt, charged
colloids are often viewed as repelling each other via a
Derjaguin-Landau-Verwey-Overbeek (DLVO) pairwise repulsion which has
the form $\Q^2\lB\exp(-\kappas r)/r$ for a pair of colloids each with
charge $\Q$ and $r$ apart \cite{Israelachvili}.  Treating such
repulsions within a mean-field approximation yields a contribution to
the free energy density which is exactly what we have found,
$\rhoprot^2/(4\rhosalt)$ \cite{Hill}.  This free energy density is
just a free energy per protein molecule of order
$\Q\rhoprot/(4\rhosalt)$.  If we assume an interface of width
$\kappas^{-1}$ at the edge of a crystal, within which there are
$(\rhoprot/\Q)\kappas^{-1}$ protein molecules per unit area of the
interface, each of which has a DLVO energy which is some fraction of
its value in the bulk, then the contribution of the DLVO repulsion to
the interfacial tension is of order $-\rhoprot^2/(\rhosalt\kappas)$.
Thus in the high salt limit our results for the contributions to the
bulk free energy density and to the interfacial tension are of the
same order as those obtained by assuming the protein molecules repel
each other via a pairwise repulsion of the DLVO form; provided we use
a mean-field approximation for the bulk and assume a protein free
energy density profile of width $\kappas^{-1}$ at the interface.  Note
that if we used a DLVO potential and assumed its contribution to
$\dsurftens$ was simply due to a missing nearest-neighbour repulsion
per molecule at the surface, then $\dsurftens=\Q^2\lB\exp(-\kappas
b)/b^3$, where $b$ is the lattice spacing.  This is rather different
from our Donnan-potential result.  Our finding that, in the high salt
limit, our results obtained via a Donnan potential approach plus the
\PB\ equation for the interface profile, can be obtained (up to
numerical prefactors) via a DLVO potential is consistent with work on
charged synthetic colloids. This work found that a DLVO pair potential
is adequate to describe the system, see for example
Ref.~\cite{Monovoukas}. By contrast, at low salt, charged colloids are
not well described by a pair potential, see Refs.~\cite{Larsen,PBW2}
and references therein.

\subsection{Predictions for lysozyme}
To make these results more concrete, we now consider a specific
example.  We choose the lysozyme/NaCl/water system, since considerable
information has been collected here.  Elsewhere it has been shown that
the crystallisation boundary in this system and the second virial
coefficient data can be fitted reasonably well with a model of hard
spheres with sticky patches \cite{PBW,CBP}.  The sticky patches are
short-ranged directional potential wells. Following Ref.~\cite{PBW} we
perform calculations for a well depth $\welldepth=7.4\,\kT$ per
patch. With attractions of this strength, and at high salt
concentration, the protein crystal coexists with a very dilute protein
solution.  We estimate the interfacial tension in this model, in the
high salt limit, to be $\surfinf\approx\welldepth/\length^2$ where
$\length^2$ is the interfacial area per protein \cite{Sear}.  Since
the dimensions of lysozyme are $4.5\times3.0\times3.0\,\nm^3 \approx
\length^3$, we estimate $\length^2 \approx 12\,\nm^2$ and consequently
$\surfinf \approx 0.6\,\kT\,\nm^{-2} = 2.5\,\milliN\,\metre^{-1}$ (by
way of comparison, this is intermediate between the interfacial
tension of a clean oil/water interface typically
$50\,\milliN\,\metre^{-1}$, and interfacial tensions that have been
measured between colloidal phases in the range
1--$20\,\microN\,\metre^{-1}$ \cite{STTL,deHL}).  To use our theory
for the electrical correction to $\surfinf$, we need the charge
density in the crystal.  Lysozyme has quite a high, $\pH$-dependent
charge, $\Q$, and for the purpose of the present calculation we use
$\Q=10$ \cite{q10} which gives $\rhoprot\approx\Q/\length^3 \approx
0.25\,\nm^{-3} = 0.4\,\Molar$.

Fig.~\ref{fig-lysotens} shows the predicted surface tension,
$\surftens = \surfinf + \dsurftens$, as a function of salt
concentration.  As might be expected, the electrical structure at the
protein crystal-solution interface starts to significantly reduce the
interfacial tension for $\rhosalt\lesssim0.4\,\Molar$. As the
physiological salt concentration is $0.15\,\Molar$, the interfacial
tension of a protein crystal {\it in vivo} is significantly reduced by
the double layer.  Our assumption that the density of protein
molecules in the solution phase coexisting with the crystal is so low
that it can be approximated by zero, is reasonable down to a salt
concentration of about $0.1\,\Molar$, with sticky patches of strength
$\welldepth\approx7.4\,\kT$ \cite{PBW}.  Below this we would need to
take account of the density of protein in the solution phase
coexisting with the crystal.

We have been unable to find any experimental determination of the
protein crystal-solution interfacial tension, but we hope that our
calculations may stimulate experimental work to confirm our results.
However, a prediction can be extracted from our theory for the
variation in protein solubility, i.e., the lysozyme concentration in
the fluid phase coexisting with the crystal, as a function of the
concentration of NaCl \cite{PoonEgel,Guo}. As noted above, at high salt
concentration the grand potential density in the protein crystal is
$\omega(-\infty) = - 2\rhosalt + \rhoprot^2 / (4\rhosalt)$. This
implies a contribution to the excess chemical potential of a protein
molecule of $\Q\rhoprot/(2\rhosalt)$.  Treating the solution phase
which coexists with the crystal as ideal (since the protein
concentration is very small) this variation of the excess chemical
potential with salt concentration yields for the solubility as a
function of salt concentration
\begin{equation}
\ln \rho_{\mathrm{sol}}(\rhosalt)=
\ln \rho_{\mathrm{sol}}(\rhosalt\to\infty)
+\frac{\Q\rhoprot}{2\rhosalt},
\label{soleq}
\end{equation}
where $\rho_{\mathrm{sol}}(\rhosalt)$ is the charge density of the
protein molecules in the dilute solution which coexists at equilibrium
with the protein crystal.  This recovers a result also found for the
high salt limit of the theory in Ref.~\cite{PBW}.  As the first term
on the right-hand side is a constant, our prediction is that the
logarithm of the solubility is inversely proportional to the salt
concentration.  This is same dependence as found in experiment
\cite{Guo} but, as noted in Ref.~\cite{PBW}, the slope predicted by
Eq.~\eqref{soleq} is close to twice the slope measured
for the experimental data.

\subsection{Comparison with surface-force apparatus measurements}
A closely related problem to that of the free energy of a single
crystal surface is that of the potential of mean force per unit area
$\df(s)$ between a pair of protein surfaces as their separation $s$ is
varied.  This is related to the interfacial tension problem since
$\df(0) = -2\surftens$ if we choose the reference state such that
$\df(\infty)=0$. Its derivative with respect to separation is the
negative of the force per unit area between two surfaces.  We expect
that the important contribution to $\df$ from the sticky patches will
be operative only when $s<\delta$, where $\delta$ is the range of the
sticky patch potential.  Previous work on the phase diagram suggests
$\delta$ is only a few percent of the protein diameter \cite{PBW,CBP}.
If $s>\delta$, $\df$ arises from the overlap of the double layers and
should be repulsive since both surfaces carry the same charge.  In
fact, we would expect $\df$ to exhibit a potential barrier at
$s\approx\delta$, of height $-2\dsurftens$ and extending out to a
distance of order $\kappas^{-1}$.

The interaction between protein surfaces is more experimentally
accessible than the interfacial tension, as it can measured via a
direct force experiment.  Sivasankar \etal\ have recently looked at
the interaction between streptavidin-covered surfaces using a surface
force apparatus \cite{SSL}. This is not quite our system as we
consider a bulk crystal whereas the experiments are on streptavidin
monolayers, but so long as the Debye screening length is smaller than
the protein diameter, the thickness of the layer over which the ion
densities vary will not be much larger than the thickness of a
monolayer of protein molecules.  Sivasankar \etal\ found that the
interaction was well described by the non-linear \PB\ equation with a
fixed effective surface charge density of order
$10\,\mC\,\metre^{-2}$.  Whilst the present theory does not calculate
the interaction between a pair of surfaces, we note that the
electrostatic potential outside the crystal is the same as would be
obtained for a surface charge density $\sigma = (4\rhosalt/\kappas)
\sinh(\phi_0/2)$, where $\phi_0$ is the potential at the surface
determined earlier (the simplest way to derive this result is via
matching the electric field strength at $z=0$).  For example, for
lysozyme at $\rhosalt = \rhoprot = 0.4\,\Molar$, we find, using this
equation, an effective surface charge density $\sigma \approx
0.06e\,\nm^{-2} \approx 10\,\mC\,\metre^{-2}$, of exactly the same
magnitude seen by Sivasankar \etal\ \cite{SSL}.

Finally we comment briefly on the case of a single protein molecule
approaching the surface of a protein crystal. In this case one would
expect a barrier of height $\approx\Q\phi_0$ to be present in the
potential of mean force, before the isolated protein encounters the
short range attraction.

\section{Conclusion}
We have estimated the contribution to the interfacial tension of a
protein crystal made by the electrical double layer at the interface.
We have used a jellium model in which the protein is replaced by a
uniform background charge density, and have solved the \PB\ equation
for the ion density profiles for a jellium half-space.  Whilst a full
solution cannot be obtained analytically, we find a good approximation
is to use the high-salt limit result in Eq.~\eqref{dshseq} with the
Donnan potential given in Eq.~\eqref{doneq}. The contribution of the
double layer is, perhaps surprisingly, {\em negative} and for a
protein not too close to its isoelectric point and of a typical size,
it amounts to roughly $10\%$ of the total interfacial tension at salt
concentrations of order $0.1\,\Molar$, see Fig.~\ref{fig-lysotens}. As
the salt concentration is increased the contribution of the double
layer {\em decreases}, due to the decreasing step in potential as the
interface is crossed and its decreasing width.  We have also briefly
considered the potential of mean force between a pair of protein
surfaces as they are brought together. The mean force is repulsive and
agrees with that found by Sivasankar \etal\ \cite{SSL}
for the repulsion between monolayers of the protein streptavidin.

The fact that the contribution of the double layer to the interfacial
tension is negative and largest at low salt concentrations suggests
that proteins will crystallise more readily at low rather than at high
concentrations of a salt such as NaCl. At lower salt concentrations
the interfacial tension will be lower, all other things being equal,
and the rate of nucleation varies as the exponential of minus the cube
of the interfacial tension. However, decreasing the concentration of
salt will make the protein crystal more soluble, it stabilises the
solution phase at the expense of the crystal \cite{PBW}.  Thus if the
salt concentration is too low the solution will be the equilibrium
phase and crystallisation will not then occur at all. Because lowering
the salt concentration lowers both the relative stability of the
crystal with respect to the solution and the interfacial tension, we
conclude that there is no clear `best' salt concentration at which to
attempt to crystallise a protein. Also, note that by salt we mean a
salt such as NaCl in the lysozyme system, in which there are no
specific ion effects such as binding of an ion to the protein.

As well as the interaction between a pair of protein surfaces, we
mention some other directions in which our theory could be extended.
Firstly, ion correlations and the effects of an inhomogeneous charge
distribution could be included.  Secondly, we have neglected excluded
volume effects and changes in dielectric properties which are
potentially important in the interior of the protein crystal.
Thirdly, we have only considered the case where the protein crystal
coexists with a dilute protein solution, such that the protein
concentration in the solution makes a negligible contribution to the
charge balance.  It has been suggested that protein crystallisation is
facilitated by a metastable liquid-liquid demixing transition
\cite{Poon,tWF,Sear02}, in which case one should certainly examine the
effect of a more concentrated protein solution phase \cite{HD}.

We would like to dedicate this paper to Jean-Pierre Hansen, on the
occasion of his 60th birthday.  Indeed, this is particularly
appropriate since it is a pleasure to record that the present
calculations were started during the recent conference to celebrate
Jean-Pierre's many scientific contributions to liquid state theory.
The work of RPS is supported by EPSRC (GR/N36981).

\bibliography{surftens}

\end{document}